\documentclass[aps,prc,twocolumn,superscriptaddress,showpacs,twoside,floatfix]{revtex4-1}
\usepackage{graphicx}% Include figure files
\usepackage{dcolumn}% Align table columns on decimal point
\usepackage{bm}% bold math
\begin{document}

\title{
Improved Empirical Parametrization of Fragmentation Cross Sections
}

\author{
K.~S\"ummerer\footnote{Electronic address: k.suemmerer@gsi.de}}

\affiliation{GSI Helmholtzzentrum f\"ur Schwerionenforschung,
Planckstr.1, D-64291 Darmstadt, Germany}
\date{\today}

\begin{abstract}
A new version is proposed for the universal empirical formula,
EPAX, which describes fragmentation cross sections in high-energy
heavy-ion reactions. The new version, EPAX 3, is shown to yield
cross sections that are in better agreement with experimental data
for the most neutron-rich fragments than the previous version. At
the same time, the very good agreement of EPAX 2 with data on the
neutron-deficient side has been largely maintained. Comparison
with measured cross sections show that the bulk of the data is
reproduced within a factor of about 2, for cross sections down to
the pico-barn range.
\end{abstract}


\maketitle

\section{Introduction}
During the last two decades, projectile fragmentation and
separation at high energies has become one of the most important
methods for the production of exotic nuclei. Production and
separation of such species can be performed at energies of around
500 to 1000 $A$ MeV at the SIS/FRS facility~\cite{frs} at GSI in
Darmstadt, Germany. At RIKEN in Wako, Japan, the RIBF facility
with the BigRIPS separator~\cite{ribf} allows to perform such
experiments at energies of up to 350 $A$ MeV. Considerably lower
energies of up to 140 $A$ MeV are attained at the A1900
separator~\cite{nscl} at NSCL in East Lansing, USA, whereas the
GANIL/LISE facility at Caen, France, operates below about 80 $A$
MeV~\cite{ganil}. In all cases, the main advantage of high-energy
projectile fragmentation as a production method lies in the fact
that very clean fragment beams can be produced; the contaminants
become less and less important with increasing energy. Moreover,
the high energies allow unambiguous event-by-event isotope
identification of the fragments produced. Again, the higher
energies help in suppressing unwanted charge states of the
fragments.

All of the above-mentioned facilities rely on good estimates of
fragmentation cross sections in order to predict the production
rates of exotic nuclei, together with ion-optical transmission
calculations with codes like {\sc MOCADI}~\cite{mocadi} or {\sc
LISE++}~\cite{lise}. One important tool to predict fragmentation
cross sections is a universal analytical formula called
EPAX~\cite{epaxv11,epaxv21}. This formula allows to calculate the
yields from fragmenting all non-fissile projectiles in the range
of projectile masses, $A_p$, between about 40 to 209. In
particular, the formula tries to take properly into account the
influence of the projectile proton or neutron excess onto the
neutron-to-proton ratio of the fragments. The cross sections are
assumed to be energy-independent, which seems to be supported by
most of the measured data, even though no systematic studies of
the bombarding-energy dependence have been performed for targets
like $^9$Be. The EPAX parametrization aims at reproducing the bulk
of the measured cross sections within a factor of two for fragment
masses down to about half the projectile mass. This latter
restriction is no severe limitation since the highest transmission
in an ion-optical projectile-fragment separator is always attained
for small mass and charge differences between projectiles and
fragments.

The previous version of the formula, EPAX 2~\cite{epaxv21}, has
proven to give rather realistic estimates of many production cross
sections of exotic nuclei. As an example, the very proton-rich
fragment $^{100}$Sn was produced with $11.2 \pm 4.7$
pb~\cite{Sch96} in $^{124}$Xe+$^9$Be (later remeasured to yield
$5.8 \pm 2.1$ pb by Straub {\it et al.}~\cite{Str10}), not far
from the EPAX 2 prediction of 7.4 pb. On the other hand, large
discrepancies with measured data were found for extremely
neutron-rich nuclei like, e.g., fragment yields from 1 $A$ GeV
$^{136}$Xe+$^9$Be, which were overestimated by EPAX by up to two
orders of magnitude~\cite{Ben08}. It is therefore desirable to try
and modify the EPAX 2 parameters in such a way that also the
yields of very neutron-rich fragments of medium- and heavy-mass
projectiles can be predicted with better accuracy.

In the following, the basic characteristics of the EPAX formula
will be reviewed.  This is followed by an explanation how the
parameters of EPAX can be obtained by fits to recent data sets.
These fits allow to get a feeling how the EPAX parameters vary
with fragment mass and with proton or neutron excess of the
projectile. The final choice of the parameters for EPAX 3 is then
presented, together with comparisons between measured data of
selected systems and the predictions of the new parametrization.

\section{The {\sc EPAX} formula}

As explained in detail in Ref.~\cite{epaxv21}, the cross section
$\sigma$ of a fragment with mass $A$ and charge $Z$ produced by
projectile fragmentation from a projectile $(A_p,Z_p)$ impinging
on a target $(A_t,Z_t)$ is written as
\begin{equation}
\sigma (A,Z) = Y_A~\sigma_Z (Z_{prob} - Z)\\
             = Y_A~n~
         exp\,(-R~|Z_{prob} - Z|^{U}).\label{eq1}
\end{equation}

The first term, $Y_A$, represents the mass yield, i.e. the sum of
all isobaric cross sections with fragment mass $A$. The second
term, $\sigma_Z$, describes the "charge dispersion", the
distribution of elemental cross sections with a given mass, $A$,
around its maximum, $Z_{prob}$. The shape of the charge dispersion
is controlled by the width parameter, $R$, and the exponent, $U$.
The factor $n = \sqrt{R/\pi}$ simply serves to normalize the
integral of the charge dispersion approximately to unity. Note
that the isobar distributions are not symmetric on the neutron-
and proton-rich side, therefore $U$ has two different values,
$U_p$ and $U_n$, on the proton- and neutron-rich side of the
valley of $\beta$-stability, respectively. In Ref.~\cite{epaxv21},
the exponent $U$ for the neutron-rich side of the isobar
distribution was chosen as $U_n=1.65$. In the present version, the
exponent on the proton-rich side is also taken as a constant,
$U_p=2.1$, close to the value of 2.0 for a Gaussian. Thus, in a
fitting procedure of the EPAX parameters to measured data, three
parameters have to be obtained for each fragment mass, $A$: An
amplitude constant, proportional to the mass yield, $Y_A$, the
centroid of the charge dispersion, $Z_{prob}$, and the width
parameter, $R$. As was shown in Ref.~\cite{epaxv21}, a fourth
parameter has to be determined for those data sets where small
cross sections of very proton-rich fragments have been measured:
the transition point, $Z_{exp}$, where the quasi-Gaussian shape of
the charge dispersion on the proton-rich side turns into an
exponential decay.

\subsection{{\sc EPAX} fits to measured data sets}

During the decade between the release of EPAX 2 and today, many
data-sets of experimental fragmentation cross-sections have been
published, some of them very comprehensive in their coverage of
fragment charge and mass numbers. In addition to the experiments
cited in the EPAX 2 paper~\cite{epaxv21} and the ones mentioned
above~\cite{Sch96,Str10,Ben08}, very comprehensive data on the
fragmentation of $^{40,48}$Ca and $^{58,64}$Ni on $^9$Be and
$^{181}$Ta targets at 140 $A$ MeV by Mocko {\it et
al.}~\cite{Moc06} turned out very useful. Tarasov {\it et
al.}~\cite{Tar10} have published a very careful study of $^{76}$Ge
fragmentation at 132 $A$ MeV, where great care was taken to
reliably model the (relatively small) transmission of the MSU
A1900 separator at this bombarding energy. Data for heavier
projectiles, all on $^9$Be targets, were measured at the FRS
separator at GSI~\cite{frs}: For $^{92}$Mo (at 500 $A$ MeV) by
Fernandez {\it et al.}~\cite{Fer05}, for $^{112}$Sn by Stolz {\it
et al.}~\cite{Sto02}, and for $^{208}$Pb by Benlliure {\it et
al.}~\cite{Kur07}, the latter all measured at 1 $A$ GeV. In
addition, a very comprehensive data set for $^{124,136}$Xe on a
$^{208}$Pb target, again at 1 $A$ GeV, has been published by
Henzlova {\it et al.}~\cite{Hen08}. A remarkable milestone was
reached by Perez and co-workers~\cite{Per11} who succeeded to
measure a rather large set of fragmentation cross sections for a
secondary beam of the extremely neutron-rich nucleus $^{132}$Sn,
obtained from $^{238}$U projectile fragmentation at 1 $A$ GeV.

In a few cases, cross sections for the same fragment have been
measured at both, medium and high bombarding energies. A very
comprehensive data set is available for the projectile $^{58}$Ni,
with data measured at 140 $A$ MeV~\cite{Moc06} and at 650 $A$
MeV~\cite{Bla94}. The experimental cross sections agree well
within their respective error bars, therefore the 140 $A$ MeV data
sets for $^{40,48}$Ca and $^{58,64}$Ni on $^9$Be of
Ref.~\cite{Moc06} where included in the fits of Eq.~(\ref{eq1}) to
isobaric cross section distributions.

The EPAX formula, Eq.~(\ref{eq1}), assumes that the free
parameters of the formula, $Y_A$, $Z_{prob}$, and $R$, vary
smoothly with fragment mass, $A$ (the parameter $U$ has a fixed
value of $U_p = 2.1$ on the proton-rich side and of $U_n = 1.65$
on the neutron-rich side). To obtain a feeling for the
$A$-dependence of the parameters, least-squares fits of Eq.~(\ref{eq1}) 
with 3
parameters were made to the cross sections of many $A$ chains of
the various systems mentioned above. Only for $^{58}$Ni, the
smallest cross sections on the proton-rich side required the
inclusion of a fourth parameter, $Z_{exp}$, which will be
explained in detail in Subsect.~\ref{twofour}. It turned out that
Eq.~(\ref{eq1}) can describe the distributions with very good
accuracy over many orders of magnitude. The significance of a good
fit is high, however, only for distributions where many data
points on both, the proton and neutron rich side are available. If
only few data points are
%----------------------------------------------------------------
\begin{figure}[h]
\vspace{-8mm}
\includegraphics[width=1.0\linewidth]{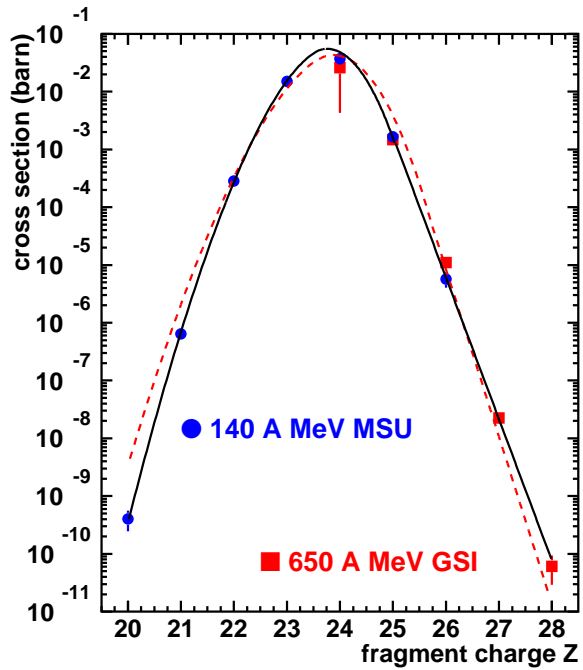}
\vspace{-8mm} \caption{(color online) Experimental cross sections
for $A=50$ isobars from $^{58}$Ni fragmentation at 140 $A$ MeV
(Ref.~\protect\cite{Moc06}, circles) and at 650 $A$ MeV
(Ref.~\protect\cite{Bla94}, squares) in a $^{9}$Be target. The
full curve represents a least-squares fit of
Eq.~(\protect\ref{eq1}) to the data. The dashed curve shows the
agreement of the EPAX 3 formula with the data. } \label{fig01}
\end{figure}
%----------------------------------------------------------------
available for a given $A$ and, in particular, if only one side of
the charge dispersion has been measured, it is difficult to obtain
a unique set of fit parameters since the parameters are strongly
correlated.

A good example is visualized in Fig.~\ref{fig01}, where
Eq.~(\ref{eq1}) has been fitted to data for 9 different isobars of
the $A=50$ mass chain, where 3 cross sections have been measured
at both energies, 140 $A$ MeV~\cite{Moc06} and at 650 $A$
MeV~\cite{Bla94}. The smallest cross sections measured amount to
sub-nanobarns. Note that on the proton-rich side the
quasi-Gaussian with an exponent of $U_p = 2.1$ turns to an
exponential slope at $Z_{exp} = 25.04$ (see
Subsect.~\ref{twofour}).

Fits of this type were made for all of the systems of
Refs.~\cite{Moc06,Bla94,Hen08,Kur07,Web94,Rei98} for those
isobaric chains where at least 4 data points are available. For
$^{208}$Pb, the data measured with a $^{nat}$Cu target by De Jong
{\it et al.}~\cite{Dej98} were combined with those measured with a
$^9$Be target by Benlliure {\it et al.}~\cite{Kur07} by scaling
the former ones with a scaling factor of 0.77 to account for the
larger size of the Cu target (see Eq.~(\ref{eq3}) below). Examples
of the fragment-mass dependence of the parameters $Y_A$,
$Z_{prob}$, and $R$ are given below in
Figs.~\ref{fig03}-\ref{fig05}.

\subsection{The {\sc EPAX 3} formula and its parameters}
\subsubsection{Mass yields}

As noted already in Ref.~\cite{epaxv11}, the mass-yield curve,
$Y_A (A)$, can to a large extent be approximated by an exponential
depending on the difference between projectile and fragment
masses, ($A_p - A$). The slope of this exponential, $P$, is a
function of the projectile mass only. The functional form of the
mass yield curve in EPAX 3 has been kept identical to the previous
version:
\begin{eqnarray}
Y_A & = & S   \cdot P \cdot exp\,(-P \cdot (A_p-A)),\label{eq2} \\
S   & = & s_1 \cdot (A_p^{1/3} + A_t^{1/3} - s_2)\;\; [{\rm barn}],\label{eq3} \\
P & = & exp\,(p_1 + p_2 \cdot A_p). \label{eq4}
\end{eqnarray}
Eq.~(\ref{eq3}) shows that the dependence of $S$ on the sum of
$A_t^{1/3}$ and $A_t^{1/3}$ and thus on the circumference of the
colliding nuclei has been kept.
%----------------------------------------------------------------
\begin{figure}[h]
\vspace{-5mm}
\includegraphics[width=1.0\linewidth]{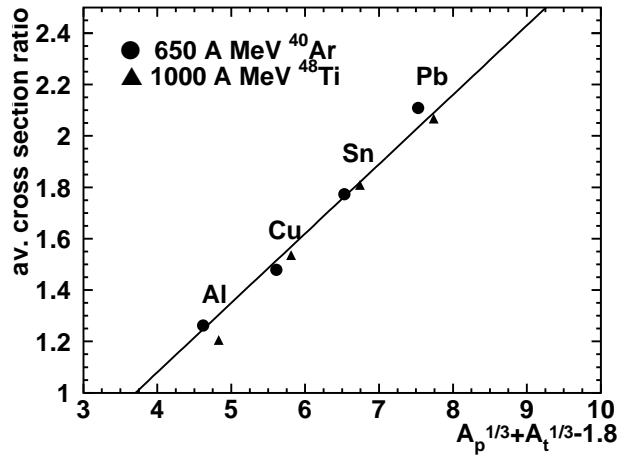}
\vspace{-5mm} \caption{Charge-changing cross sections of
relativistic $^{40}$Ar and $^{48}$Ti beams interacting with the
targets indicated in the figure, relative to those with C targets.
The ratios were averaged over all charge changes measured in
Ref.~\protect\cite{Zei08}. The straight line represents the EPAX
description, Eq.~(\protect\ref{eq3}).}\label{fig02}
\end{figure}
%----------------------------------------------------------------
As the only change the overlap factor, $s_2$, has been slightly
reduced from 2.38 to 1.8 fm. Results from a recent paper on
charge-changing cross sections induced by 650 $A$ MeV $^{40}$Ar
and 1 $A$ GeV $^{48}$Ti ions by Zeitlin {\it et al.}~\cite{Zei08}
confirm that Eq.~(\ref{eq3}) is a good description of the
geometrical scaling factor. Fig.~\ref{fig02} shows the ratios of
charge-changing cross sections for Al, Cu, Sn, and Pb targets,
relative to those of a carbon target, and averaged over all charge
changes measured in Ref.~\cite{Zei08}. The x-axis corresponds to
the scaling variable in Eq.~(\ref{eq3}), $(A_p^{1/3} + A_t^{1/3} -
s_2)$; the straight line represents Eq.~(\ref{eq3}).

\noindent In Ref.~\cite{epaxv21} it was shown that the mass yield
$Y_A$ has to be increased near the projectile (see Eq.~(13) of
Ref.~\cite{epaxv21}). The same formula was kept for the new
Version 3:
\begin{equation}
Y_A'  = Y_A \cdot [y_2 \cdot A_p \cdot (x-y_1)^2] \label{eq5}
\end{equation}
for $x \ge y_1$; $x$ is the ratio of fragment mass to projectile
mass, $x = A/A_p$. The numerical values of all constants can be
found in Table~\ref{tab_const}.
%************** table ******************************************************
\begin{table}[b]
\caption{Constants used in the new EPAX 3 formula. }
\begin{tabular}{lld}
\hline\hline
Parameter & Constant & Value \\
\hline
Scaling factor $S$              & $s_1$     & 0.27     \\
                                & $s_2$     & 1.80    \\
Mass yield slope $P$            & $p_1$     & -1.731    \\
                                & $p_2$     & -0.01399  \\
Mass yield correction factor    & $y_1$     & 0.75     \\
                                & $y_2$     & 0.10      \\

$Z_{prob}$ shift $\Delta$       & $d_1$     & -1.087    \\
                                & $d_2$     & 3.047 \cdot 10^{-2}\\
                                & $d_3$     & 2.135 \cdot 10^{-4}\\
                                & $d_4$     & 71.35     \\
                                & $d_5$     & -25.0     \\
                                & $d_6$     & 0.80      \\
n-rich memory effect $\Delta^n_m$ & $n_1$   & 0.40      \\
                                  & $n_2$   & 0.60      \\
p-rich memory effect $\Delta^p_m$ & $q_1$   & -10.25    \\
                                  & $q_2$   &  10.25    \\
n-rich slope  $U_n$             & $U_n$     & 1.65      \\
p-rich slope  $U_p$             & $U_p$     & 2.10      \\
Width parameter $R$             & $r_0$     & 2.78      \\
                                & $r_1$     & -0.015    \\
                                & $r_2$     & 3.20 \cdot 10^{-5}\\
                                & $r_3$     & 0.0412    \\
                                & $r_4$     & 0.124     \\
                                & $r_5$     & 30.0      \\
                                & $r_6$     & 0.85      \\
Transition point to             & $l_1$     & 1.20      \\
 exponential slope              & $l_2$     & 0.647     \\
Downscale factor                & $b_1$     & 2.3 \cdot 10^{-3} \\
                                & $b_2$     & 2.4       \\
\hline \hline
\end{tabular}
\label{tab_const}
\end{table}
%************** table ******************************************************

\begin{figure}[hbt]
%\vspace{3cm}
\includegraphics[width=1.00\linewidth]{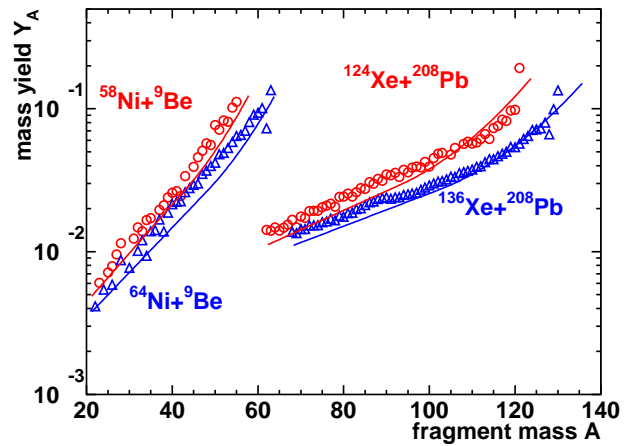}
\vspace{-5mm} \caption{(color online) Mass yields, $Y_A(A)$, for
$^{58,64}$Ni+$^9$Be and $^{124,136}$Xe+$^{208}$Pb. The open
symbols represent the numerical values of $Y_A$ derived from fits
of Eq.~(\protect\ref{eq1}) to the experimental data of
Refs.~\protect\cite{Moc06,Hen08}. The full lines have been
calculated from Eqs.~(\ref{eq2}-\ref{eq5})}\label{fig03}
\end{figure}

Typical examples of mass yields are shown in Fig.~\ref{fig03}. The
data points represent the numerical values of $Y_A$ as obtained
from fitting Eq.~(\ref{eq1}) to isobaric chains of cross sections
of Refs.~\cite{Moc06,Hen08}. One can clearly see that the slopes,
$P$, are much steeper for lighter projectiles like $^{58,64}$Ni
than for heavier ones like $^{124,136}$Xe, corroborating the $A_p$
dependence of $P$, Eq.~(\ref{eq4}). Eq.~(\ref{eq5}) has been
chosen such that the upward bend of $Y_A$ towards the projectile
mass scales with $A_p$ and thus has only a minor effect on the
$^{58,64}$Ni mass yield distributions. The slope differences
between each of the two Ni and Xe isotope pairs are entirely due
to the $A_p$-dependence of $P$.
%----------------------------------------------------------------
%----------------------------------------------------------------
\subsubsection{Centroid $Z_{prob}$}
%----------------------------------------------------------------
\begin{figure}[htb]
\begin{center}
\vspace{-5mm}
\includegraphics[width=1.00\linewidth]{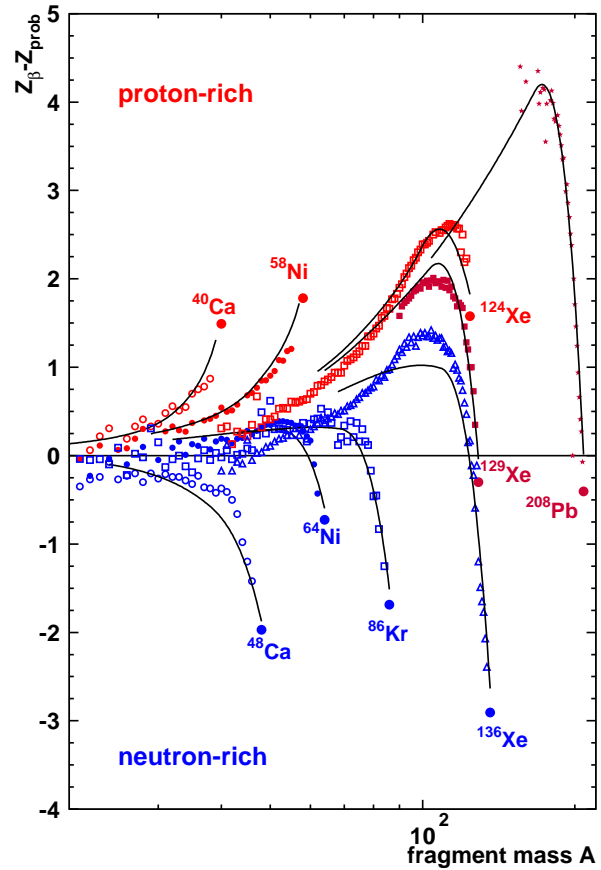}
\end{center}
\vspace{-7mm} \caption{(color online) Loci of the peak cross
sections, $Z_{prob}(A)$, as a function of fragment mass, for
various projectiles as indicated in the Figure. For clarity, the
value of $Z_{\beta}(A)$ has been subtracted at each mass number
$A$. The data points represent numerical values of $(Z_{\beta}-
Z_{prob})$ obtained by fitting Eq.~(\protect\ref{eq1}) to isobaric
distributions for the systems indicated. Positive (negative)
ordinate values correspond to proton-rich (neutron-rich) nuclei
relative to $Z_{\beta}$. The large dots indicate where the
respective projectile is located in this graph. The lines
represent the EPAX predictions according to
Eqs.~(6-11).}\label{fig04}
\end{figure}
%----------------------------------------------------------------
As mentioned above, the charge dispersion is mainly characterized
by its centroid, $Z_{prob}$, and its width parameter, $R$. As in
EPAX Versions 1 and 2, the loci of maximum cross section,
$Z_{prob}(A)$, have been parameterized relative to the valley of
$\beta$-stability,
\begin{equation}
Z_{prob} = Z_{\beta} + \Delta + \Delta_m. \label{eq6}
\end{equation}

$Z_{\beta}$ is approximated by the smooth function
\begin{equation}
Z_{\beta}=A/(1.98 + 0.0155 \cdot A^{2/3}). \label{eq7}
\end{equation}
$\Delta$ is found to be a linear function of the fragment mass,
$A$, for heavy fragments ($A \geq d_4$), and is extrapolated
quadratically to zero:
\begin{equation}
\Delta=\left\{ \begin{array}{ll}
        d_1 + d_2 \cdot A  &
          \;\;\mbox{if $A \geq d_4$} \\
        d_3 \cdot A^{2}
        &\;\;\mbox{if $A < d_4$} \end{array}
        \right.\label{eq8}
\end{equation}
The sum $Z_{\beta} + \Delta$ defines the ``residue corridor'', the
locus in the $A - Z$ plane where the fragments from all
projectiles located near $\beta$-stability end up after long
evaporation chains. This corridor is close to the line
Charity~\cite{Cha98} has found from studying the residue
distributions after evaporation from highly excited compound
nuclei (see dashed line in Fig.~4 and Eqs.~(6,7) in
Ref.~\cite{Cha98}). This points to the fact that the fragment
distributions far away from the projectiles are largely controlled
by evaporation from equilibrated pre-fragments.

It is obvious that close to the projectile, $Z_{prob}$ has to move
away from the corridor and approach $Z_p$, i.e., for projectiles
located on the line of $\beta$-stability, $\Delta$ has to approach
zero. That has been achieved by the following modification
\begin{equation}
\Delta' = \Delta \cdot [1 + d_5 \cdot (x-d_6)^2] \label{eq9}
\end{equation}
if $x>d_6$.

For projectiles on the proton- or neutron-rich side of the valley
of $\beta$-stability, an additional correction of $Z_{prob}$ is
necessary, which accounts for the ``memory effect'', the gradually
vanishing proton or neutron excess of the fragment reflecting the
proton or neutron excess of the projectile, $(Z_p - Z_{\beta p})$,
where $Z_{\beta p}$ is the value of Eq.~(\ref{eq7}) for the
projectile, $A = A_p$. This additional correction, termed
$\Delta_m$ in Eq.~(\ref{eq6}), should reflect the full excess at
the projectile mass and should vanish gradually with increasing
mass loss from the projectile. The functional forms have been kept
as in the previous version,
\begin{equation}
\Delta^n_m = (n_1 \cdot x^2 + n_2 \cdot x^4) \cdot (Z_p-Z_{\beta
p}) \label{eq10}
\end{equation}
for neutron-rich projectiles, and
\begin{equation}
\Delta^p_m  = exp\;(q_1 + q_2 \cdot x) \cdot (Z_p-Z_{\beta p})
\label{eq11}
\end{equation}
for proton-rich projectiles. In both cases, the choice of the
parameters guarantees the limiting properties mentioned above.

Fig.~\ref{fig04} visualizes how the combined effects of $\Delta$,
$\Delta'$, and $\Delta_m$ compare to numerical values of
$(Z_{\beta}-Z_{prob})$ derived from fits of Eq.~(\ref{eq1}) to the
experimental data sets of Refs.~\cite{Moc06,Hen08,Web94,Rei98}.
The zero line corresponds to the line of $\beta$-stability
according to Eq.~(\ref{eq7}). The full lines have been calculated
using Eqs.~(\ref{eq6}-\ref{eq11}). The agreement with the fitted
data points is rather good close to the respective projectile
masses; further away, one notes bad agreement in particular for
$^{136}$Xe.

\subsubsection{Width parameter R}

%----------------------------------------------------------------
\begin{figure}[htb]
\begin{center}
\vspace{-8mm}
\includegraphics[width=1.00\linewidth]{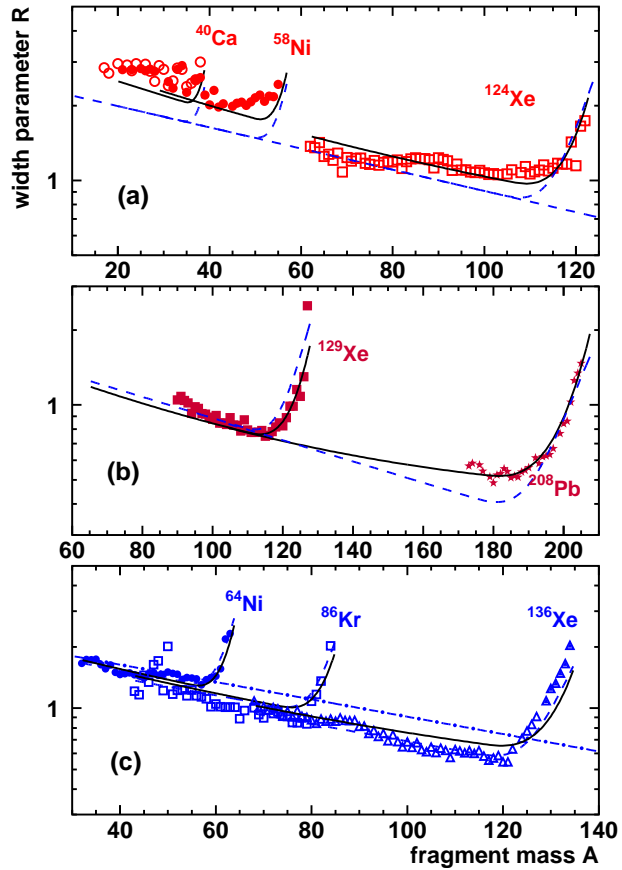}
\end{center}
\vspace{-8mm} \caption{(color online) (a) Fragment-mass dependence
of the width parameter, $R(A)$, for proton-rich projectiles. The
symbols represent numerical values of $R$ obtained by fitting
Eq.~(\ref{eq1}) to the systems indicated. The dashed curves denote
the old EPAX 2 parametrization, while the new Version 3
(Eqs.~(\ref{eq12}-\ref{eq15})) is represented by the full curves.
(b) The same for projectiles near the line of $\beta$-stability.
Note the reduced widths (larger $R$ values) for
$^{208}$Pb-fragments according to EPAX 3. (c) The same for
neutron-rich projectiles. Since here the dashed curves are almost
invisible, the dash-dotted line indicates the EPAX 2 curve for
stable and proton-rich projectiles. } \label{fig05}
\end{figure}
%----------------------------------------------------------------
Similar to the parameter $Z_{prob}$ just discussed, the width
parameter, $R$, is to first order a function of fragment mass
only, irrespective of the projectile. In Ref.~\cite{epaxv21} it
was found that the experimental $R$-values can be approximated by
a simple exponential (see Eq.~(8) of Ref.~\cite{epaxv21}). As
visualized in the following section, this yields too wide charge
dispersions for very heavy nuclei, such that now a parabolic
dependence in log scale has been chosen:
\begin{equation}
R = R_0 \cdot exp\,(r_1 \cdot A + r_2 \cdot A^2) \label{eq12}
\end{equation}
It was already noted in Ref.~\cite{epaxv21} that the charge
dispersions from neutron-rich projectiles are a bit wider than
those from proton-rich or $\beta$-stable ones. In the present
version of EPAX, this has been coded in the following form: the
parameter $R_0$ in Eq.~(\ref{eq12}) depends differently on the
proton and the neutron excess of the projectile:
\begin{eqnarray}
R^n_0 & = & r_0 \cdot exp\,[r_3 \cdot (Z_p - Z_{\beta p})] \label{eq13}\\
R^p_0 & = & r_0 \cdot exp\,[r_4 \cdot (Z_p - Z_{\beta p})]
\label{eq14}
\end{eqnarray}
It is again obvious that close to the projectile the widths of the
charge dispersions should shrink since the smaller number of
nucleons removed from the projectile also reduces the variance of
the mass loss. Therefore, the following modification has been
introduced near the projectile, i.e. for $x>r_6$:
\begin{equation}
R' = R \cdot exp\,[r_5 \cdot \sqrt{A_p} \cdot (x - r_6)^3].
\label{eq15}
\end{equation}
The $\sqrt{A_p}$-dependence of the pre-factor in the exponent
guarantees a small effect for light and a stronger one for heavy
projectiles.

Fig.~\ref{fig05} compares the width parameters calculated from
Eqs.~(\ref{eq12}-\ref{eq15}) to numerical values of $R$ obtained
by fitting Eq.~(\ref{eq1}) to experimental cross sections of the
various systems. The upper part, (a), shows this comparison for
proton-rich projectiles ($^{40}$Ca, $^{58}$Ni, and $^{124}$Xe).
The EPAX 3 parametrization (full lines) leads to more narrow
charge dispersions than the EPAX 2 one (dashed lines).
Fig.~\ref{fig05}b has been plotted for the projectiles $^{129}$Xe
and $^{208}$Pb, both located close to the line of
$\beta$-stability. The more narrow widths compared to the previous
parametrization~\cite{epaxv21} for $^{208}$Pb are clearly visible.
The lower panel, (c), displays $R(A)$ for three neutron-rich
projectiles, $^{64}$Ni, $^{86}$Kr, and $^{136}$Xe. The
parametrization according to EPAX 2 coincides with the data points
and is hardly visible; there is not much difference to the Version
3 parametrization (full lines). The dash-dotted line indicates the
EPAX 2 exponential for stable or proton-rich projectiles; it is
obvious that neutron-rich projectiles lead to considerably wider
distributions.

\subsubsection{\label{twofour} Modifications for very proton-rich fragments}

A major modification of EPAX 2 compared to EPAX 1 consisted in the
introduction of an exponential rather than a quasi-Gaussian shape
of the charge distribution beyond a certain transition point on
the proton-rich side~\cite{epaxv21}. This was coded by calculating
the derivative of the logarithm of the cross section
(Eq.~(\ref{eq1})):
\begin{equation}
\frac{dF}{dZ} = \frac{d(log(\sigma))}{dZ} \approx \frac{-2 \cdot
R}{ln(10)} \cdot (Z-Z_{prob}) \label{eq16}
\end{equation}
The transition point to the exponential slope, $Z_{exp}$, has been
calculated for the proton-rich side as a function of the fragment
mass $A$ according to
\begin{equation}
Z_{exp}(A) = Z_{prob}(A) + {\frac{dF}{dZ}\Bigg\vert}_A  \cdot
\frac{ln(10)} {2 \cdot R(A)} \label{eq17}
\end{equation}
From $Z_{exp}$ on, the slope is exponential with the same gradient
as Eq.~(\ref{eq1}) at this point. The empirical parametrization
for the transition slope proposed in Ref.~\cite{epaxv21},
\begin{equation}
\frac{dF}{dZ} = l_1 + l_2 \cdot (A/2)^{0.3} \label{eq18}
\end{equation}
not only yields perfect agreement with very proton-rich fragments
from $^{58}$Ni, but provides the same quality also for fragments
from $^{124}$Xe, e.g. for $^{100}$Sn, for which EPAX 2 calculates
a cross section of 7.4 pb, whereas Straub {\it et al.} measure
$5.8 \pm 2.1$ pb~\cite{Str10}. This parametrization has therefore
been kept in EPAX 3, the $^{100}$Sn cross section from this new
version now amounts to 6.9 pb.

\subsubsection{Modifications for very neutron-rich fragments}

It has been noted by several authors (e.g. by Benlliure {\it et
al.}~\cite{Ben08,Kur07}) that EPAX 2 overestimates cross sections
of very neutron rich fragments from, e.g., $^{136}$Xe or
$^{208}$Pb by up to two orders of magnitude. After many failed
attempts to modify the EPAX parameters in such a way that the
discrepancies were removed, it became clear that only a
``brute-force solution'' could solve this problem. This means that
all cross sections from neutron-rich projectiles calculated by
EPAX 3 with Eqs.~(1-18) above should be downscaled by a factor
that depends on both, the neutron excess of the projectile and the
fragment. This ``brute-force factor'' can be written as
\begin{eqnarray}
f_{bf}  =  10^{ [B  \cdot (Z_{\beta}(A) - Z + b_2)^3]}\label{eq19} \\
B = - b_1 \cdot |Z_p - Z_{\beta p}|\label{eq20}
\end{eqnarray}
where $Z_{\beta}$ and $Z_{\beta p}$ are the values of
Eq.~(\ref{eq7}) for fragment mass $A$ and projectile mass $A_p$,
respectively.  It applies to fragments where $(Z_{\beta}-Z) >
(Z_p-Z_{\beta p} + b_2)$. The effect of this factor is quite
dramatic for the most neutron-rich fragments of heavy projectiles
as will be shown below.

\section{Comparisons with experimental cross sections}

\subsection{Projectiles close to $\beta$-stability}

\begin{figure*}[hb]
\begin{minipage}[c]{0.70\linewidth}
\includegraphics[width=1.00\textwidth]{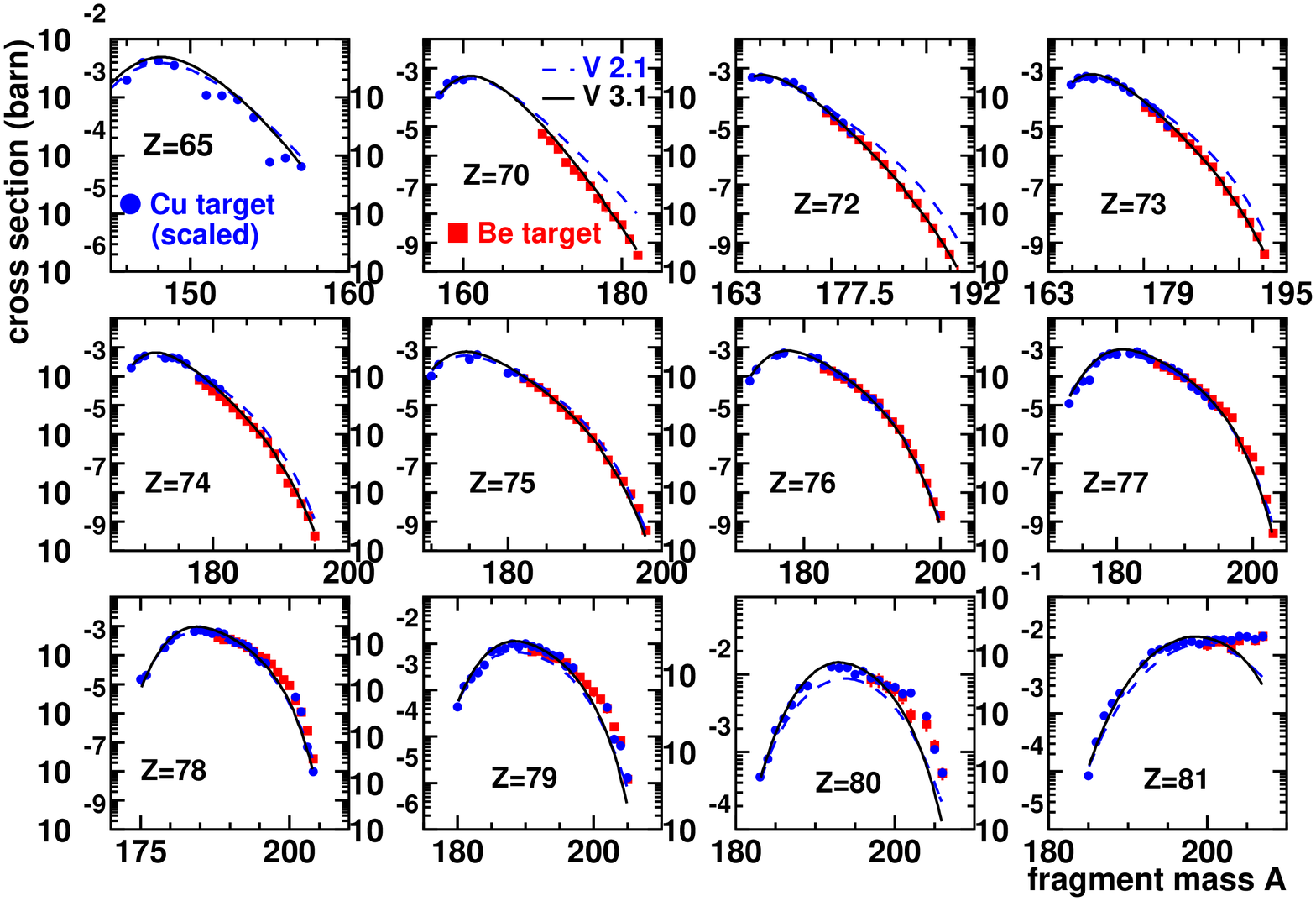}
\end{minipage}
\begin{minipage}[c]{0.25\linewidth}
\vspace{-5mm} \caption{(color online) Isotope distributions from 1
$A$ GeV $^{208}$Pb fragmentation in a $^{nat}$Cu (circles,
Ref.~\protect\cite{Dej98}) and a $^9$Be (squares,
Ref.~\protect\cite{Kur07}) target. The Cu-target data were scaled
by a factor of 0.77 according to Eq.~(\protect\ref{eq3}). The full
line represents the new EPAX 3 parametrization, whereas the dashed
line shows the old EPAX 2 version. } \label{fig06}
\end{minipage}
\end{figure*}

An important testing ground for the quality of EPAX 3 predictions
are the fragmentation cross sections of $^{208}$Pb, one of the few
heavy projectiles to explore the yet poorly studied region in the
chart of nuclides ``south'' and ``south-east'' of this nucleus.
Data were measured at 1 $A$ GeV bombarding energy on
$^{nat}$Cu~\cite{Dej98} and $^9$Be~\cite{Kur07} targets. For the
following comparison (Fig.~\ref{fig06}) the cross sections of
Ref.~\cite{Dej98} were
%----------------------------------------------------------------
%----------------------------------------------------------------
scaled by a factor of 0.78 according to Eq.~(\ref{eq3}). Both data
sets fit excellently to each other and are well reproduced over a
large range of fragment masses by EPAX 3. In particular, the
predictions for neutron-rich fragments farther away from the
projectile are now much closer to the experimental data than those
of EPAX 2. Small discrepancies remain for the most neutron-rich Tl
and Hg isotopes.
%----------------------------------------------------------------
\begin{figure*}[htb]
\begin{minipage}[c]{0.70\linewidth}
\includegraphics[width=\textwidth]{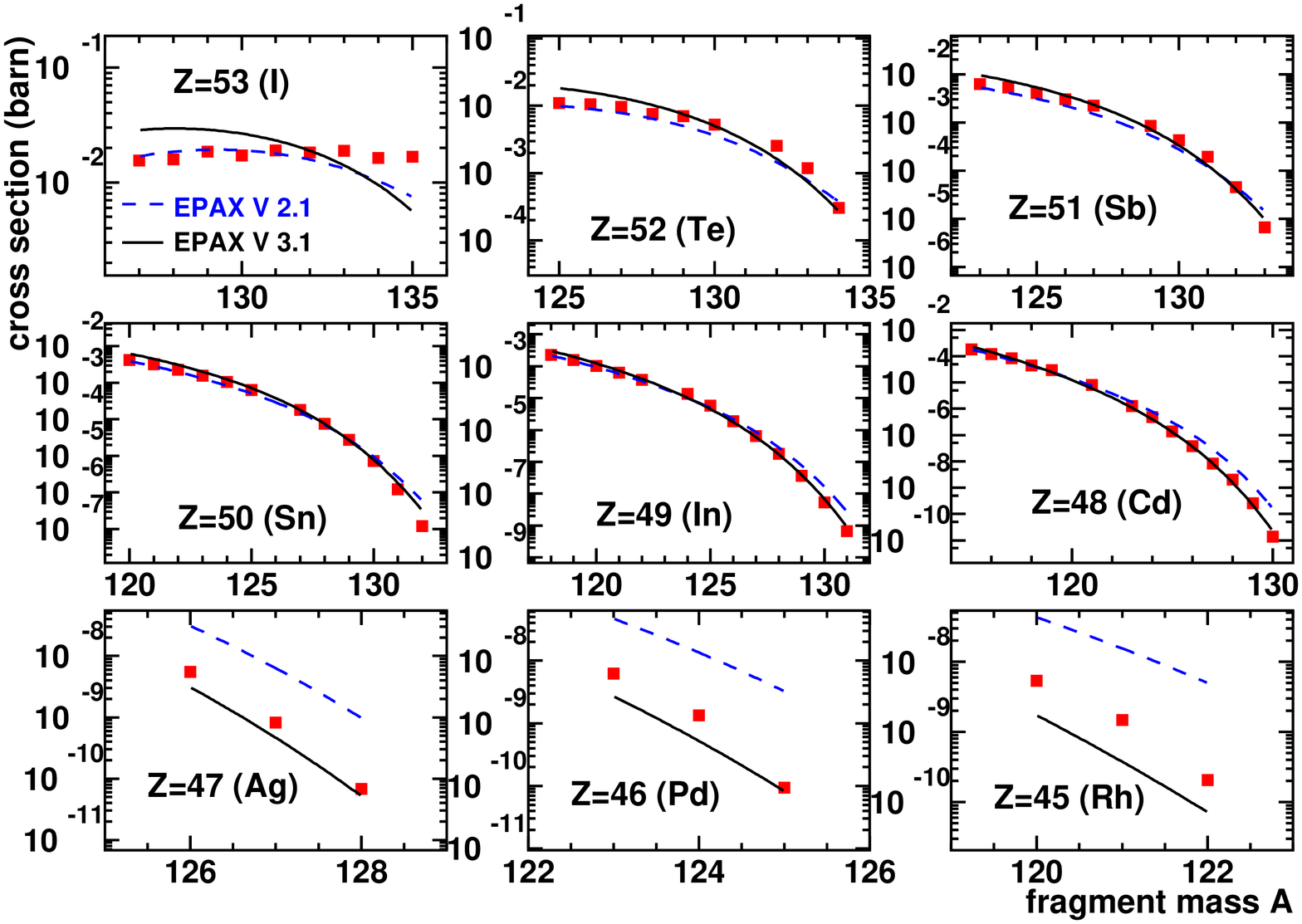}
\end{minipage}
\begin{minipage}[c]{0.25\linewidth}
\caption{(color online) Same as Fig.~\protect\ref{fig06} but for
the system $^{136}$Xe+$^9$Be~\protect\cite{Ben08}. EPAX 3 behaves
worse than EPAX 2 very close to the projectile but fits better to
the measured data points for most of the very neutron-rich
fragments. } \label{fig07}
\end{minipage}
\end{figure*}
%----------------------------------------------------------------

\subsection{Fragmentation of neutron-rich projectiles}

It was mentioned in the Introduction that one of the most severe
deficiencies of EPAX 2 was its overprediction of cross sections of
very neutron-rich fragments, e.g. those from
$^{136}$Xe+$^9$Be~\cite{Ben08}. It is therefore interesting to see
the effect of the new parameters and in particular that of the
``brute-force'' downscale factor (Eqs.~(\ref{eq19},\ref{eq20})).
Some isotope distributions from this system are shown in
Fig.~\ref{fig07}. The flat isotope distribution for $Z=53$ is not
so well reproduced, but for the bulk of the cross sections of very
neutron-rich fragments EPAX 3 fits better than the previous
version. For the few data points measured for $Z=45$, the new
description seems even to underpredict slightly the data.

\begin{figure}[hbt]
\begin{center}
\includegraphics[width=1.0\linewidth]{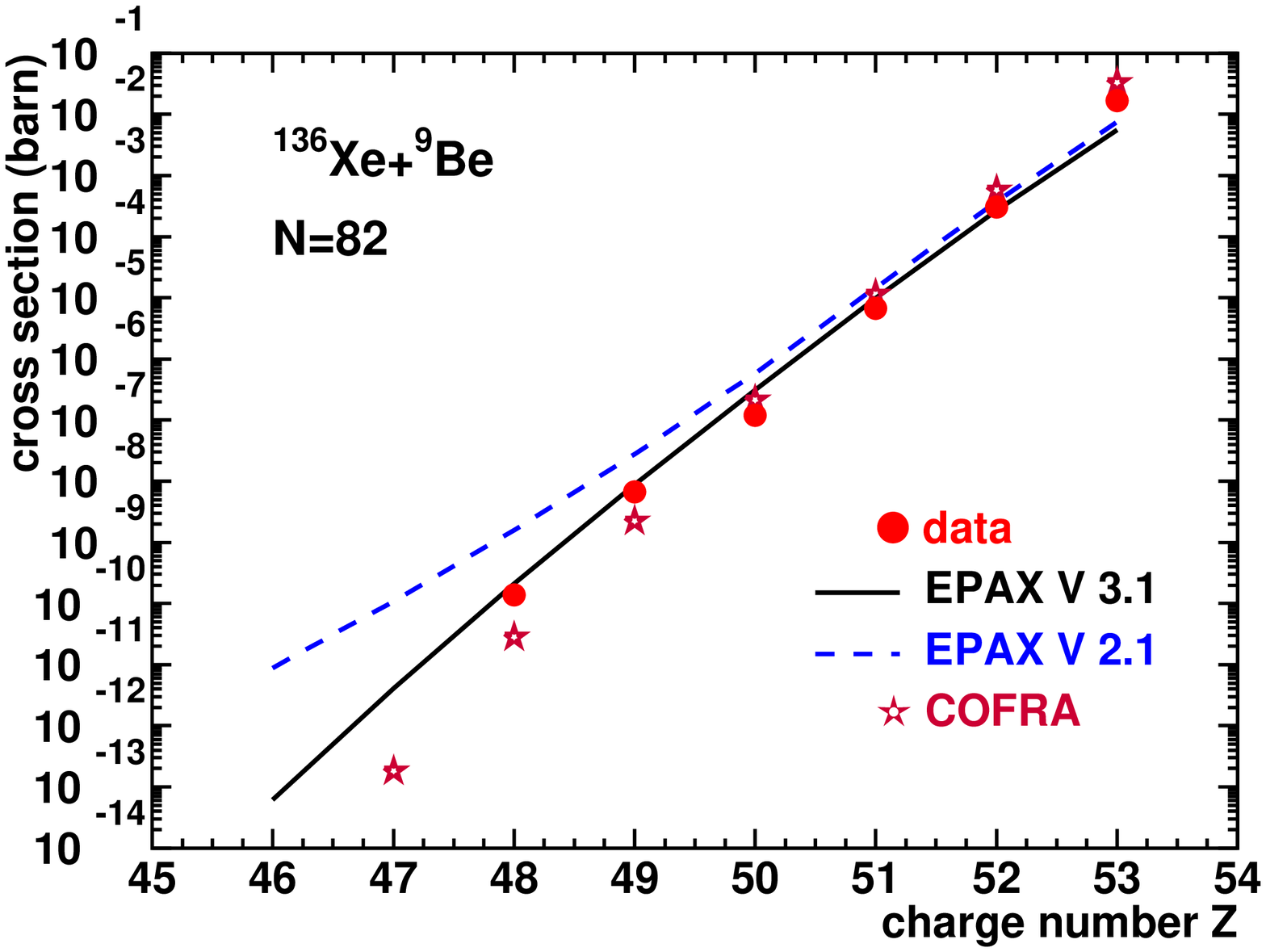}
\vspace{-8mm} \caption{(color online) Cross sections of
proton-loss channels for 1 $A$ GeV $^{136}$Xe + $^9$Be~(dots,
Ref.~\protect\cite{Ben08}) in comparison with EPAX 3 (full line),
EPAX 2 (dashed curve), and with the physical ``cold
fragmentation'' model {\sc COFRA}~\protect\cite{cofra} (stars). }
\label{fig08}
\end{center}
\end{figure}

Similar conclusions can be drawn if one plots the cross sections
for the proton-loss channels with $N=82$ from $^{136}$Xe+$^9$Be
(Fig.~\ref{fig08}). Benlliure {\it et al.} have noted in
Ref.~\cite{Ben08} that EPAX 2 yields too flat a slope for the
cross sections of the $N=82$ isotones.
%----------------------------------------------------------------
%----------------------------------------------------------------
Fig.~\ref{fig08} now demonstrates that the new parametrization
gives a slope in very good agreement with measured data, despite a
discrepancy, however, at $Z=53$.

Fig.~\ref{fig08} compares EPAX 3 also to the predictions of a
physical model, the ``cold fragmentation'' model
COFRA~\cite{cofra}. This model, a simplified analytical version of
the abrasion-ablation model which is valid in regions where only
neutron evaporation plays a role, has been shown to reproduce the
cross sections of many neutron-rich fragments very well (see,
e.g., Refs.~\cite{Ben08,Kur07,Per11}). The same is true for the
$Z=50-53$ data points in Fig.~\ref{fig08}. The model tends to
slightly underpredict, however, the lowest-$Z$ data points, where
EPAX 3 seems to follow the data a bit better.

\subsection{Fragmentation of proton-rich projectiles}

%----------------------------------------------------------------
\begin{figure}[hbt]
\begin{center}
%\vspace{3cm}
\includegraphics[width=1.0\linewidth]{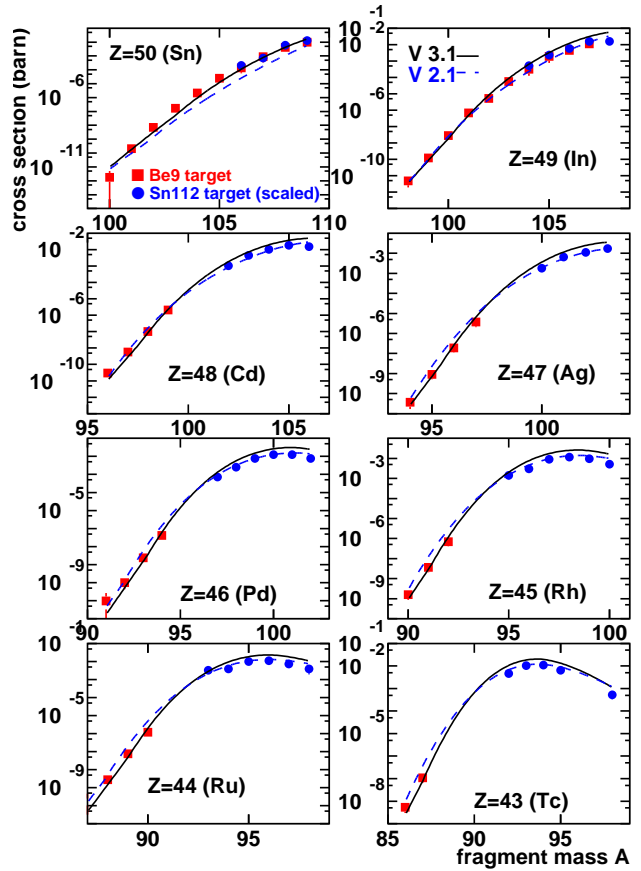}
\vspace{-8mm} \caption{(color online) EPAX 3 (full lines) and EPAX
2 (dashed lines) predictions for production cross sections from
$^{112}$Sn+$^9$Be, in comparison with data on a $^9$Be
(Ref.~\protect\cite{Sto02}, squares) and a $^{112}$Sn
(Ref.~\protect\cite{Foe11}, circles) target measured at 1 $A$ GeV.
The latter data points have been scaled by a factor of 0.65. }
\label{fig09}
\end{center}
\end{figure}
%----------------------------------------------------------------
Contrary to the neutron-rich fragments discussed in the previous
subsection, cross sections for even extremely proton-rich
fragments like $^{48}$Ni or $^{100}$Sn were already predicted with
surprising accuracy by EPAX 2. Care was undertaken in the present
study not to deteriorate the excellent predictive power of EPAX on
the proton-rich side. Fig.~\ref{fig09} shows that this has been
achieved to a large extent. The figure shows measured cross
sections for $^{112}$Sn+$^9$Be~\cite{Sto02}, complemented by data
for less exotic fragments from the system
$^{112}$Sn+$^{112}$Sn~\cite{Foe11}, both data sets obtained at 1
$A$ GeV. The Sn-target data were scaled by a factor of 0.65 (see
Eq.~(\ref{eq3})) to match the Be-target data. Fig.~\ref{fig09}
confirms that both data sets fit together smoothly and that EPAX 3
describes the entire distributions rather well, very similar to
EPAX 2. As a benchmark test, one can check the cross section of
$^{100}$Sn from $^{112}$Sn+$^9$Be fragmentation: The agreement
with the measured value of $1.8^{+2.3}_{-1.3}$ pb~\cite{Sto02} has
deteriorated a bit, but not much (Version 3 predicts 7.9 pb,
compared to 6.6 pb for Version 2). Another benchmark is provided
by the fragment $^{48}$Ni produced from $^{58}$Ni+$^{58}$Ni at 160
$A$ MeV with a cross section of $100 \pm 30$ fb~\cite{Pom12}. This
is to be compared with an EPAX 3 prediction of 20 fb, EPAX 2
predicted 57 fb.

\subsection{Range of validity of the {\sc EPAX} formula}

As mentioned in the Introduction, the EPAX parametrization
contains no bombarding-energy dependent terms, following the
concept of ``limiting fragmentation'', which assumes that the
cross sections for the formation of fragments rather close to the
projectile are independent of bombarding energy if the latter is
sufficiently above the Fermi energy in nuclei ($\approx \, 40 \, A
\, MeV$). This concept is difficult to prove since only few
systems have been investigated where identical fragments were
produced at several different bombarding energies. One system is
$^{58}$Ni+$^9$Be which was studied at 140 and 650 $A$
MeV~\cite{Moc06,Bla94}. An example how well results from these two
studies fit to each other was shown above in Fig.~\ref{fig01};
other mass chains exhibit similar agreement. Another study also
points to the fact that around 140 $A$ MeV energy independence is
largely achieved: the reaction $^{76}$Ge+$^9$Be at 132 $A$ MeV
studied by Tarasov {\it et al.}~\cite{Tar10}. Most of their cross
sections of very neutron-rich fragments in the range $23 \leq Z
\leq 17$ are quantitatively reproduced by EPAX 3.

On the other hand, a comparison of two $^{86}$Kr-induced data
sets, one measured at 500 $A$ MeV~\cite{Web94} and the other at 64
$A$ MeV~\cite{Moc07}, do not yield identical results. EPAX fits to
the latter yield systematically narrower width parameters, $R$.
Moreover, the simple geometrical scaling law for two different
targets, which seems to work at high energies (see
Figs.~\ref{fig06},\ref{fig09}), breaks down as visualized in
Fig.~11 of Ref.~\cite{Moc07}. In agreement with Mocko {\it et al.}
we conclude that an energy of 64 $A$ MeV is below the range of
validity of the EPAX formula.

\section{Summary}

Some modifications of the EPAX 2 parameters were shown to improve
the predictive power of the EPAX formula in regions where too
large cross sections were calculated, in particular for very
neutron-rich fragments from medium-mass and heavy projectiles. In
addition to a modified parameter set, a ``downscale factor'' had
to be introduced for neutron-rich projectiles that depends on the
neutron-excess of the projectile and scales with the
neutron-excess of the fragment. The resulting EPAX 3 formula
reproduces with rather good accuracy the bulk of the measured
fragmentation cross sections as long as the incident projectile
energy is larger than 130 - 140 $A$ MeV. Systematic discrepancies
were found for $^{86}$Kr projectiles at 64 $A$ MeV. The
improvement of EPAX 3 compared to the previous version is
confirmed by the much better agreement with the physical
``cold-fragmentation'' model {\sc COFRA}, which was found to
reproduce well measured cross sections for the most neutron-rich
fragments from $^{136}$Xe and $^{208}$Pb.

The program code of the EPAX formula can be downloaded from the
GSI Document Server~\cite{GSI12}.

\section*{Acknowledgments}

Sincere thanks are due to many colleagues who have provided
numerical data of their results, sometimes prior to publication.
Among them are in particular H.~Alvarez Pol, J.~Benlliure,
D.~Henzlova, A.~Kelic, J.~Kurcewicz, T.~Kurtukian Nieto, D.~Perez
Loureiro, M.~Mocko, A.~Stolz, O.~Tarasov, and B.~Tsang. The author
acknowledges gratefully the kind hospitality of the Grupo
Experimental de Nucleos y Particulas at the Universidad de
Santiago de Compostela where part of this work was completed.

%----------------------------------------------------------------
%
\end{document}